\documentclass[
10pt, 
a4paper, 
oneside, 
headinclude,footinclude, 
BCOR5mm, 
]{scrartcl}

\usepackage{geometry}
\usepackage[
nochapters, 
beramono, 
eulermath,
pdfspacing, 
dottedtoc 
]{classicthesis} 

\usepackage{arsclassica} 

\usepackage[T1]{fontenc} 

\usepackage[utf8]{inputenc} 

\usepackage{graphicx} 
\graphicspath{{Figures/}} 

\usepackage{enumitem} 

\usepackage{lipsum} 

\usepackage{subfig} 

\usepackage{amsmath,amssymb,amsthm} 

\usepackage{arydshln} 

\usepackage{varioref} 

\usepackage{float}

\theoremstyle{definition} 

\theoremstyle{plain} 

\theoremstyle{remark} 

\hypersetup{
colorlinks=true, breaklinks=true, bookmarks=true,bookmarksnumbered,
urlcolor=webbrown, linkcolor=RoyalBlue, citecolor=webgreen, 
pdftitle={}, 
pdfauthor={\textcopyright}, 
pdfsubject={}, 
pdfkeywords={}, 
pdfcreator={pdfLaTeX}, 
pdfproducer={LaTeX with hyperref and ClassicThesis} 
} 

\usepackage{geometry}
\setheadwidth{\textwidth}

\usepackage[style=numeric, autocite=superscript]{biblatex}
\usepackage{caption}
\usepackage[export]{adjustbox} 
\usepackage[htt]{hyphenat}

\theoremstyle{definition}

\numberwithin{equation}{section}

\title{\normalfont{Variable selection in linear mixed model meta-regression with suspected interaction effects - How can tree-based methods help?}} 

\author{\small{Jan-Bernd Igelmann\textsuperscript{*1}, Paula Lorenz\textsuperscript{1}, Markus Pauly\textsuperscript{1,2}}} 

\date{Preprint version - not peer reviewed\\
\vspace{0.2 cm}
\today} 

\begin{document}


\renewcommand{\sectionmark}[1]{\markright{\spacedlowsmallcaps{#1}}} 
\lehead{\mbox{\llap{\small\thepage\kern1em\color{halfgray} \vline}\color{halfgray}\hspace{0.5em}\rightmark\hfil}} 

\pagestyle{scrheadings} 


\maketitle 

\setcounter{tocdepth}{2} 




\abstract{\noindent Detecting interaction effects (IEs) in meta-regression is challenging, especially when few studies are available and many plausible interactions are considered. In many meta-analyses, interpretability is essential, which limits the use of complex machine learning methods. Tree-based approaches offer a potentially useful compromise, but their role in meta-regression with random effects is not yet well understood. This paper examines how traditional linear and tree-based methods can support variable selection for IEs in random effects meta-regression.
We compare test-based and information-criterion-based linear selection procedures with meta-CART approaches. These include single trees and tree ensembles, which combine single meta-CARTs into a stability selection ensemble based on bootstrapped data. All methods are evaluated using a real-world meta-analytic dataset and a simulation study. The data-generating process assumes linear IEs, complemented by settings with simple non-linear interactions.
Our results show that under strictly linear interactions, linear selection methods perform as expected and achieve superior performance for IE detection. Tree-based methods are more conservative when the number of studies is small, but become competitive as sample size increases, particularly the stability-selected variants. When IEs deviate from strict linearity, Wald-type approaches can deteriorate, whereas criterion-based methods maintain a relatively stable performance.
Tree-based methods, particularly stabilized versions of random-effects meta-CART, provide a robust alternative when a sufficient number of observations is available.
They could thus be used for pre-selection and sensitivity analyses, guarding against more complex data structures.
Additionally, selection frequency patterns from meta-CART tree ensembles can help to reveal structural patterns in the data in an exploratory way.}

\section{Introduction}

\subsection{Meta-regression and the difficulty of variable selection}
A methodological key challenge in meta-analysis is handling possible \textit{heterogeneity}.\autocite{thompson1999Explaining, higgins2002Statistical, thompson2002How}
If the source of heterogeneity is unknown and no further study information is available, the most common solution is to fit a random effects model.
A non-systematic heterogeneity structure, often following a normal distribution, is implicitly assumed.\autocite{jackson2018When}
In cases, where additional study information (covariates) are at hand, these can be used to model the heterogeneity structure and possibly detect its sources. 
This is usually done by applying regression approaches and thus is referred to as \textit{meta-regression} \autocite{berkey1995Randomeffects, thompson2002How, baker2009Understanding, tipton2019History, welz2020Simulation}. Linear regression approaches including an additive term for modeling the influence of the covariates are the gold-standard. This form gives a type of stability, makes results comparable and provides a structure for testing the significance of effects. Additional flexibility is offered by the possibility to include interaction and higher order terms.

A possible approach to include relevant covariates is to rely on external knowledge. 
This means pre-defining covariates before discovering any data patterns (\textit{a priori identification})\autocite{baker2009Understanding}. 
This is not always possible due to the lack of reliable knowledge and reduces the chance of detecting additional structures within the data. 
\citeauthor*{baker2009Understanding} stresses the retrospective character that a meta-analysis has.\autocite{baker2009Understanding}
Thus, alternatively, influential covariates can be determined in an exploratory way. \autocite{lissa2020Small, baker2009Understanding}
Accordingly, the chance of detecting sources of heterogeneity increases but so does the chance of identifying influential factors erroneously. This chance of so-called \textit{spurious findings} is particularly large due to the heterogeneous character of meta-analyses but also due to the often small number of studies \autocite{thompson2002How, higgins2004Controlling}. 
Stable variable selection methods are needed that are able to detect true structures (low Type~II error) while minimizing the chance of spurious findings (low Type~I error).

\subsection{Variable selection including interaction effects and the marginality principle}
Regression results, and consequently variable selection results, can be particularly unstable and prone to overfitting when the ratio $p/k$, with $p$ denoting the number of covariates and $k$ the number of studies, is small. \autocite{austin2015Number,geissbuhler2021Most} 
This is already the case when only main effects~(MEs) and no interaction effects~(IEs) are considered but becomes more severe when IEs are taken into account.
When IEs are taken into account, the number of potential parameters can become large. For example, for $p = 6$, a model including all MEs and all pairwise IEs comprises
\textit{1~intercept + 6~MEs + 15~IEs = 22} parameters. 
A common rule of thumb, which must be interpreted cautiously depending on the setting, is to allow roughly one parameter per ten studies \autocite{peduzzi1996Simulation}. In our example with $p = 6$, this would imply 220 studies which is unrealistic for most meta-regressions. The recent review by  \citeauthor*{geissbuhler2021Most}\autocite{geissbuhler2021Most} reported a median of $k =23$ studies per meta-analysis (IQR 13 to 41 studies, with $N = 81$ analyzed meta-studies). Moreover, the Cochrane Collaboration's recommendation that meta-regressions should not be performed with fewer than ten studies further indicates that many applied investigations operate close to this threshold\autocite{cochrane_handbook_10_11_4}. 
A consequence is that the possibility of existing IEs often is ignored and thus results are misinterpreted. An example for this is given by a re-analysis of the meta-analysis of \citeauthor*{kimmoun2021Temporal}\autocite{kimmoun2021Temporal}.
Including interaction terms shows that the originally stated time trend could be confounded by an ignored IE between the time and the average patient age \autocite{knop2023Consequences}. Furthermore, methodological research shows that it is not always necessary to ignore IEs for reasons of estimation accuracy. If the correct IEs are specified, robust estimation and inference methods can deliver reliable results, also in a meta-regression \autocite{thurow2024Robust}.

A central aspect in selecting IEs is the \textit{marginality principle} \autocite{nelder1977Reformulation}.
In the additive linear model considered here, including an IE while omitting its associated MEs violates the hierarchical structure of the model. Since the IE is generally not orthogonal to the MEs, their omission can lead to biased parameter estimates and an incorrect allocation of explained variance.
As a consequence, classical $F$-tests for assessing whether an IE equals zero are no longer valid as the residual space is misspecified and the resulting degrees of freedom are incorrect. Beyond these statistical issues, excluding the MEs also undermines the substantive interpretation of the interaction, as IEs describe how the effect of one variable depends on the level of another, a concept that is ill-defined when the corresponding main effects are absent. 
Yet, recent theoretical discussions argue that the distinction between an IE and its MEs can be somewhat arbitrary and dependent on the chosen measurement scale, unless there is a clear substantive hierarchy among variables\autocite{morris2023Marginality}.
Under such transformations, the definition of “main” versus “interaction” effects may shift. This model dependence extends beyond measurement scale to the choice of model class itself\autocite{hand1994Deconstructing}: In tree-based models, for instance, interactions are represented implicitly through successive (conditional) splits rather than through explicit interaction terms. This perspective is already reflected in the name of one of the earliest regression tree methods, \textit{Automatic Interaction Detection}\autocite{morgan1963Problems}.
In meta-analytic settings, however, variables are often very conventional and naturally defined (e.g., study year, average patient age, or categorical variables such as study country) and interpretability is an essential part of a meta-analysis. 
In such contexts, the marginality principle provides both conceptual clarity and methodological stability. Moreover, by enforcing a hierarchical structure, it helps to control model dimensionality. We therefore adopt this principle and apply it throughout this paper.

Overall, even when adhering to the marginality principle, the number of candidate models grows rapidly with $p$. Specifically, there are $N(p)$ admissible models, where
\begin{equation}
     N(p) = \sum_{k=0}^{p} \binom{p}{k} \, 2^{\binom{k}{2}}.
\end{equation}  
The largest model contains $ 1 + (p(p+1))/2$ parameters.
Thus, classical variable selection procedures, typically comparing all possible models up to the full model, can hardly succeed. Either the model fails to converge, or the resulting estimates are highly unstable, particularly in meta-analytic settings that include random effects and a small number of observations. Therefore, in practice, reduced solutions are required when IEs should not be ignored.

\subsection{Variable selection approaches}

From an analyst’s perspective, two general strategies are available for addressing the problems described above. Either a pre-selection of IEs is performed, followed by a classical variable selection procedure, or single-step procedures are applied that accommodate both the marginality principle and a large number of parameters.
The pre-selection strategy reduces the maximum number of parameters to be estimated. This approach is consistent with general recommendations that potential IEs should be considered carefully \emph{a priori} \autocite{baker2009Understanding} and is particularly compatible with enforcing the marginality principle \autocite{nelder1977Reformulation}, as also advocated in practical guidelines, see, e.g. p.~139 in the book by \citeauthor*{weisberg2014applied}\autocite{weisberg2014applied}.
However, two-step variable selection procedures can be problematic, as they risk overoptimism due to sequential, data-driven decisions based on the same data set and may fail to detect important dependencies among predictors. 

Within these two general strategies, we focus on methods that can accommodate the large number of potential IEs while respecting the marginality principle. These methods can either be interpreted as single-step selection approaches or be used to generate a reduced set of candidate IEs, to which a second selection step may be applied. The latter step, however, is not considered explicitly in this work.

Among the considered methods, we examine procedures that are strictly based on linear model structures. These include approaches relying on statistical hypothesis testing as well as procedures based on information criteria. We consider both univariate and multivariate testing strategies. Univariate tests are frequently used in meta-analytic applications, primarily due to their straightforward implementation and their flexible handling of predictors with missing observations \autocite{tipton2019Current, cinar2021Using}. 
With respect to information-criterion-based methods, we focus on the AICc, the small-sample corrected version of the AIC \autocite{hurvich1991bias}, and the BIC \autocite{schwarz1978Estimating}. We also include tree-based variable selection approaches, which have been characterized to be particularly promising for detecting IEs \autocite{lissa2017MetaForest, dusseldorp2014Combinations}.

\subsection{Tree-based approaches}
A meta-analysis is an exploratory or retrospective tool used to combine results obtained from studies investigating the same topic or other comparable data sources\autocite{baker2009Understanding}. The overall goal is to discover connections and new information based on often limited amounts of data. Most modern machine-learning techniques are not well suited for these descriptive
tasks, as they tend to function as black-box algorithms whose value lies primarily in predictive performance\autocite{breiman2001statistical}. However, several tree-based approaches have recently been developed specifically for meta-analysis\autocite{dusseldorp2014Combinations, li2017MetaCART, li2019Flexible, li2020Multiple, lissa2017MetaForest, lissa2020Small}. Various authors have, for example, acknowledged\autocite{tipton2019History} or applied\autocite{bull2018Interventions, parr2021Using} types of these methods. Random forest-based methods, such as \emph{MetaForest}\autocite{lissa2020Small}, can assess the importance of variables and use this information for variable selection via comparing \textit{feature importance}. Simpler classification and regression tree models, such as \emph{meta-CART}\autocite{li2020Multiple}, construct decision trees in which the splits define subgroups that explain as much heterogeneity as possible. Such models have been shown to be effective when the data-generating model (DGM) indeed follows a tree-like structure\autocite{li2019Flexible, li2020Multiple}. 
While doing so, tree-based methods implicitly perform a form of variable selection. Variables that appear along a tree can therefore be suspected to have an effect that may also be captured by a classical linear model. In particular, effects that occur along a single branch of a tree may be interpreted as a form of interaction. Owing to these properties, tree-based methods can be used as variable selection procedures for classical linear models with suspected IEs, either as a pre-selection step or even as a one-step modeling approach. The idea of detecting IEs in a meta-analysis via tree-based methods has been described as promising \autocite{lissa2017MetaForest, dusseldorp2014Combinations}. A key potential opportunity of these approaches is that purely linear IEs can be detected reasonably well, while being more robust than linear methods when the underlying effects deviate from strict linearity. An additional, albeit secondary, advantage of meta-CART is that it can naturally handle missing covariate values through so-called surrogate splits, which constitutes a particularly important benefit in meta-regression settings, where missing study-level information is common, as already discussed in the context of univariate analyses.

In this paper, we investigate the corresponding performance in a simulation study. The focus is on (generalized) linear mixed models with suspected IEs. In contrast to existing studies on meta-CARTs \autocite{li2017MetaCART, li2019Flexible}, we refrain from using DGMs based solely on binary splits, and in contrast to other studies \autocite{lissa2020Small}, we specify IEs in a classically interpretable parametric form. Furthermore, we do not aim to compare predictive performance.

In the simulation study, we include linear variable selection methods and a meta-CART approach \autocite{li2020Multiple}. Ensembles of trees generally lead to more stable results than single trees. As no established meta-CART ensembles exist that are completely suitable for the selection task, we additionally consider a stabilized bootstrapped tree-based approach. This serves as a representative for various tree ensemble methods that could be adapted to this setting and follows ideas similar to those underlying MetaForest\autocite{lissa2020Small}. The emphasis of our simulation framework is on evaluating how both the classical linear procedures and the tree-based procedures perform in variable selection with the target structure being an IE in a classical random effects meta-regression model.

\section{Statistical methods}
After introducing the linear meta-analytical model specifications (Section~\ref{sec:model}) and the corresponding estimation procedures (Section~\ref{sec:est}), we present the analyzed variable selection methods. These methods are grouped into two categories: those that rely on a linear model structure (Section~\ref{sec:lin_var_meth}) and those based on tree-structured models (Section~\ref{sec:tree_var_meth}).  
We included approaches that can either be used as a pre-selection procedure for IEs or (in an adapted version) could also be applied to full model selection involving IEs and comply with the marginality principle. As discussed previously, a comprehensive selection step with substantially fewer parameters could be conducted after the pre-selection stage. However, this second step is beyond the scope of the present analysis.

For all linear variable selection methods, we present the random effects specification as the more general case rather than the fixed effect (or common effect) model. The fixed effect model can be obtained as a special case by setting $\tau^2 = 0$. 
We emphasize that the choice between fixed and random effects models should always be justified on substantive grounds. In practice, however, it is often unrealistic to assume the absence of residual heterogeneity in data sets where covariates are explicitly used to explain between-study variability. For this reason, we rely on random effects modeling in the following analyses and include a DGM setting with $\tau^2 = 0$ in the simulation study. 
Furthermore, since tree-based methods constitute the main focus of the evaluation and fixed and random effect Meta-CARTs are based on distinct algorithms, we consider both specifications in parallel.

An alternative, and commonly used approach for variable selection, that is not included in this paper, is the LASSO regularization\autocite{tibshirani1996Regression, friedrich2023regularization}.
A valuable Bayesian version for meta-regression is available in the \texttt{R} package \texttt{pema}\autocite{vanlissa2023Selecting}, which offers a principled way to incorporate shrinkage priors into the random effects meta-regression framework.
However, its shrinkage priors are specified independently for each coefficient, which does not directly align with the marginality structure required for IEs in our setting.
Combining this with the marginality structure is beyond the scope of the present paper and needs further investigation.

\subsection{Mixed linear models for meta-regression}

\label{sec:model}

We consider a meta-analysis including $k \in \mathbb{N}$ independent studies, each reporting an effect size estimate~$y_i \in \mathbb{R}$ and a sample variance $v_i > 0$ ($i = 1,\dots, k$). Since primary studies often do not report uncertainties of the sampling variance and involve a sufficiently large number of individuals,
it is commonly assumed to be known \autocite{dersimonian1986Metaanalysis}. This is also fundamental to the maximum likelihood or weighted least squares (WLS) estimation and inference as defined below. 

Different types of effect sizes $y_i$ are accepted. For non-normal effect sizes, a prior transformation to the real-valued scale is assumed, such as Fisher's $r$-to-$z$ transformation for Pearson correlation coefficients\autocite{fisher1915Frequency, welz2022Fisher}.

Study characteristics reported by each study are denoted as $p \in \mathbb{N}$ covariates $x_{i,1}, \ldots, x_{i,p}$. These covariates may represent basic study characteristics, such as the year of publication, or aggregated information, such as the mean patient age. Hence, the measurement scale can vary. Metric variables are assumed to be standardized, meaning that their original mean is subtracted and the result is divided by the corresponding standard deviation. This can improve numerical stability and is essential for several variable selection procedures\autocite{hastie2009elements}. Categorical variables can be encoded using dummy variables, with one binary indicator for each factor level except the reference category. 
This increases the number of candidate predictors. When some classes have very low frequencies, it can be preferable to merge them. 
In addition, a block-wise testing approach must be adapted when performing variable selection based on hypothesis testing (see Section \ref{sec:testing}). 
The most commonly used regression approach in this setting is the random effects meta-regression model given by
\begin{equation}
\label{eq:REMA}
    y_i = \tilde{\beta}_0 + \tilde{\beta}_1 x_{i,1} + \ldots + \tilde{\beta}_p x_{i,p} + u_i + \epsilon_i,
\end{equation}
with regression coefficient $\tilde{\beta}_i\in\mathbb{R}$, independent random effects terms $u_i \sim \mathcal{N}(0,\tau^2)$, heterogeneity parameter $\tau^2 > 0$, and independent within-study errors $\epsilon_i \sim \mathcal{N}(0,v_i)$. Random effects and within-study errors are also assumed to be independent. 
If there is no evidence of heterogeneity, a fixed effect model is obtained by setting $\tau^2 = 0$\autocite{borenstein2021Introduction}. 
However, this model ignores the potential presence of IEs and includes all available covariates, even if some have no actual influence. 
Therefore, we adjust our assumed model to
\begin{equation}
\label{eq:mod_IE}
    y_i = \beta_0 
    + \sum_{j \in \boldsymbol{J}} \beta_j x_{i,j}
    + \sum_{(l,l') \in \boldsymbol{L}} \beta_{l,l'} x_{i,l} x_{i,l'}
    + u_i + \epsilon_i,
\end{equation}
where $\boldsymbol{J} \subseteq \{1, \ldots, p\}$ denotes the subset of covariates assumed to be influential. In addition, potential IEs are included through $\beta_{l,l'}$ for $(l,l') \in \boldsymbol{L}$, with $\boldsymbol{L} \subset \boldsymbol{J}^2$ and $l \neq l'$, following the marginality principle.

In practice, more (or fewer) covariates may affect the response. Unobserved influences are absorbed by $u_i$ or the residual error $\epsilon_i$. Such omitted variables may, however, violate the normality assumption and introduce bias.
Consequently, both the estimated coefficients and the identified influential covariates and IEs are subject to uncertainty. Errors thus arise not only from coefficient estimation but also from the selection of $\boldsymbol{J}$ and $\boldsymbol{L}$.
During the modeling process, the model is fitted for various choices of $\boldsymbol{J}$ and $\boldsymbol{L}$. 

Practical challenges include \textit{missing information} and \textit{aggregation bias}. In meta-analysis, some moderators are not reported by all studies. As stated above, we assume that all included covariates are available for each of the $k$ studies. In practice, this implies that either covariates were selected to ensure complete availability, a complete-case set of studies was used, or missing values were imputed. 
Aggregation bias (or ecological fallacy) arises when covariates are aggregated over study participants\autocite{baker2009Understanding}. While this does not affect true study-level characteristics such as publication year, it may be problematic for variables like patient age. In many applications, these inaccuracies are negligible, and we therefore assume that all covariates are measured without error and are free from aggregation bias.

\subsection{Estimation}
\label{sec:est}

If we combine the coefficients from the random effects meta-regression model in Equation~\eqref{eq:mod_IE} into a single vector $\boldsymbol{\beta} \in \mathbb{R}^{m}$ with $m = 1 + \# \boldsymbol{J} + \# \boldsymbol{L}$ (intercept + number of MEs + number of IEs), Equation~\eqref{eq:mod_IE} can be rewritten in matrix notation as
\begin{equation}
    \boldsymbol{y} = \boldsymbol{X}\boldsymbol{\beta} + \boldsymbol{u} + \boldsymbol{\epsilon}
\end{equation}
with $\boldsymbol{y} \in \mathbb{R}^k$, $\boldsymbol{X} \in \mathbb{R}^{k \times m}$, $\boldsymbol{u} \sim \mathcal{N}_k(\boldsymbol{0},\tau^2 \boldsymbol{I}_{k \times k})$ and $\boldsymbol{\epsilon} \sim \mathcal{N}_k(\boldsymbol{0}, \boldsymbol{V})$. Here $\boldsymbol{I}_{k \times k}$ denotes the $(k\times k)$ identity matrix and $\boldsymbol{V} \in \mathbb{R}^{k\times k}$ is a diagonal matrix with the sampling variances $v_i$ on its main diagonal.

$\text{Var}(\boldsymbol{y})^{-1}$ is defined via the weight matrix $\boldsymbol{W} = \text{diag}(w_1, \ldots, w_k)$ and $w_i = (v_i + \tau^2)^{-1}$. Based on this, the log-likelihood function is given as
\begin{equation}
     \ell(\boldsymbol{\beta}, \tau^2) = -\frac{k}{2}\ln(2\pi) - \frac{1}{2} \ln \left| \boldsymbol{W}^{-1} \right| - \frac{1}{2} \left(\boldsymbol{y} - \boldsymbol{X} \boldsymbol{\beta}\right)' \boldsymbol{W} \left(\boldsymbol{y} - \boldsymbol{X} \boldsymbol{\beta}\right).
     \label{eq:ll}
\end{equation}

For a fixed value of $\tau^2$, $\ell$ is \emph{maximized by the maximum-likelihood estimator}\autocite{cinar2021Using} $\hat{\boldsymbol{\beta}}$ (or equivalently the weighted-least-squares estimator), which is defined as
\begin{equation}
    \hat{\boldsymbol{\beta}} = \left( \boldsymbol{X}' \boldsymbol{W} \boldsymbol{X}\right)^{-1} \boldsymbol{X} \boldsymbol{W} \boldsymbol{y}.
\end{equation}
The selection of an estimator $\hat{\tau}^2$ is treated separately. We employ the approximately unbiased \emph{restricted maximum-likelihood estimator} (REML)\autocite{viechtbauer2005Bias, cinar2021Using}. The restricted likelihood is used only for estimating $\tau^2$ and therefore not discussed in detail.

For estimating the covariance structure of $\hat{\boldsymbol{\beta}}$, denoted as $\boldsymbol{\Sigma}$, we use the standard HKSJ estimator, introduced independently by \citeauthor{knapp2003Improved}\autocite{knapp2003Improved} and \citeauthor{sidik2005note}\autocite{sidik2005note}. It is defined as
\begin{equation}
    \hat{\boldsymbol{\Sigma}} = s^2 \left(\boldsymbol{X}' \hat{\boldsymbol{W}} \boldsymbol{X}\right)^{-1},
\end{equation}
with $s^2 = (k - m)^{-1}(\boldsymbol{y}' \hat{\boldsymbol{W}} \boldsymbol{P} \boldsymbol{y})$ and $\boldsymbol{P} = \boldsymbol{I}_{k \times k} - \boldsymbol{X}( \boldsymbol{X}' \hat{\boldsymbol{W}} \boldsymbol{X})^{-1} \boldsymbol{X}' \hat{\boldsymbol{W}}$.
We note, however, that robust alternatives exist, in particular heteroscedasticity-consistent HC1 estimator  \autocite{mackinnon1985Heteroskedasticityconsistent}, which has shown competitive performance in simulation studies involving IEs\autocite{thurow2024Robust}.

\subsection{Classical variable selection for models with linear underlying structure}
\label{sec:lin_var_meth}
We consider both univariate and multivariate testing approaches and different information criteria.
\subsubsection{Univariate Testing}
\label{sec:testing}
The \emph{univariate} testing approaches are based on $\alpha$-level hypothesis tests of the form $H_0: \beta_j = 0$ with $\alpha \in (0,1)$; typically $\alpha = 0.05$. In practice, different types of implementations exist. We assess a Wald-type test approach via
\begin{equation}
    T_j  = \frac{\hat{\beta}_j}{\sqrt{\hat{\boldsymbol{\Sigma}}_{j,j}}} \; \overset{H_0}{\sim} \; \boldsymbol{t}_{\nu},
    \label{eq:t_dist}
\end{equation}
where $\boldsymbol{t}_{\nu}$ is the $t$-distribution with $\nu = (k - m)$ degrees of freedom\autocite{knapp2003Improved}. Consequently, an effect of the corresponding covariate is assumed when $\mid T_j \mid \geq t_{\nu, 1-\alpha/2}$, where $t_{\nu, 1-\alpha/2}$ denotes the $1-\alpha/2$ quantile of $\boldsymbol{t}_{\nu}$. The notation for parameters referring to IEs is equivalent. 

The univariate approach investigates each parameter separately. A model is fitted containing only the covariate under investigation, and the corresponding hypothesis is tested. Due to the marginality principle, which is essential for constructing valid model structures, a slight modification is required for IEs. For testing $\beta_{l,l'}$, the respective model must also include $\beta_l$ and $\beta_{l'}$, although only $\beta_{l,l'}$ is tested. Because the focus lies on variable selection, adjustments for multiple testing are omitted, which is common practice in meta-regression \autocite{cinar2021Using}.

\subsubsection{Multivariate testing}
\textit{Multivariate} testing procedures can rely on different strategies. While there exists more exhaustive approaches suitable for less possible parameters, we rely on forward selection. The forward procedure begins testing all possible univariate models and adds the covariate with the smallest $p$-value (based on Equation~\eqref{eq:t_dist}). This is repeated until no new variables are significant or when all variables are included.
The classical procedure has to be adapted for IEs due to the marginality principle. Here, a ME is automatically added when a corresponding IE enters the model. 

We focus on forward selection rather than backward selection or hybrid procedures.
This avoids fitting overly large models or even the full model with an excessive number of parameters.
Instead, it starts with smaller models, which tend to be more stable.
Problems may arise when the forward testing procedure is still ongoing, but the number of parameters becomes too large, potentially leading to unstable models. To address this, we define two additional stopping criteria.
To estimate the models, we use the \texttt{rma} function from the \texttt{metafor}\autocite{viechtbauer2010Conducting} package with REML estimation and a maximum of 1,000 iterations. 
If a candidate model fails to converge within these iterations, no further moderators are added. 
Moreover, no additional moderators are included if any of the standard errors of the standardized parameter estimates (excluding the intercept) exceeds 100, that is, if for any parameter $\sqrt{\hat{\Sigma}_{j,j}} > 100$.

In principle, multivariate testing is beneficial compared to the univariate approach, as it can more appropriately account for dependencies among covariates.
However, in the meta-analytic context many studies do not report all covariates. Here, the univariate approach has the advantage that neither studies nor covariates need to be excluded during the selection process\autocite{cinar2021Using}.
Depending on the structure of the missing data\autocite{vanbuuren2018Flexible}, this approach can nevertheless lead to issues, as a model that appears locally optimal within a subset may no longer be optimal once all covariates or all studies are considered.

\subsubsection{Information criteria}

Information criteria balance model fit and complexity by penalizing overfitting, as the likelihood increases monotonically with the number of parameters. For the present comparison, we use the unrestricted likelihood from Equation~\eqref{eq:ll}, because the restricted likelihood is, by design, not comparable across models with different fixed effect structures and is therefore unsuitable for variable selection\autocite{verbeke2009Linear}. Although recent studies suggest that REML often yields similar results\autocite{cinar2021Using, gurka2006Selecting}, we rely on the unrestricted likelihood because it is the methodologically appropriate choice. 

Two commonly used criteria are the \emph{Akaike Information Criterion} (AIC) \autocite{akaike1974New} which tends to select more fixed effects, and the \emph{Bayesian Information Criterion} (BIC)\autocite{schwarz1978Estimating}, which tends to select fewer. For the AIC, a small-sample corrected version exists that selects fewer fixed effects in small samples. 
It is defined as
\begin{equation}
    \text{AICc} = -2\ell + 2(m + 1)\left( \frac{k^\ast}{k^\ast- (m+1) -1 }\right),
\end{equation}
with $k^\ast = \max(k, m + 3)$\autocite{hurvich1991bias, cinar2021Using}. Since we focus on small-sample applications, we include the AICc and the classical BIC.  

Although information criteria are frequently employed for variable selection, they are conceptually model comparison criteria rather than genuine variable selection procedures. Variable selection arises only indirectly through the imposed model search strategy, not from the criteria themselves. The theoretically optimal approach is to exhaustively fit all possible models and select the one with the lowest criterion value. Forward and backward selection strategies can be applied, but they are usually not recommended because the optimal model can easily be missed. 
In our case, similar to the multivariate testing approach, we rely on forward selection. In a first step, the criterion is fitted for all univariate models and the variable with the lowest value is added to the model. This variable is added to all further univariate models, and, again, the variable from the model with the lowest criterion value is added. This procedure stops when no new model has a lower criterion value than the one from the previous step. The termination rules are the same as for testing. For models that contain IE parameters, both MEs are also included (marginality principle).

An alternative is model averaging, where models are weighted according to their criterion values\autocite{cinar2021Using, burnham2004multimodel}. However, methodological difficulties and non-comparable model structures arise once IEs are included. We therefore do not apply model averaging in our analyses.

\subsection{Tree-based approaches in meta-regression}
\label{sec:tree_var_meth}
Below, we explain both, CART and stabilized trees, and their application to meta-analysis.

\subsubsection{Meta-CART}
To introduce tree-based methods for variable selection in meta-regression, we begin with \emph{classification and regression trees} (CARTs). A dedicated implementation for meta-analytic applications is provided by the \emph{meta-CART} approach, which is available in the \texttt{R} package \texttt{metacart}\autocite{R_metacart}. This approach extends the interpretable CART algorithm\autocite{breiman1984classification} to the meta-analytic setting by accommodating between-study heterogeneity. In particular, meta-CART incorporates study-level weights and uses heterogeneity-reduction criteria that account for random effects. 
The splits within the CARTs are created by maximizing between-subgroup heterogeneity assessed via the $Q$-value, which is commonly used in meta-analysis \autocite{borenstein2021Introduction}. Exemplified for the first split, the weighted mean for two possible subgroups $\mathcal{T}_1 \subset \{1, \ldots, k\}$ and $\mathcal{T}_2 = \{1, \ldots, k\} \setminus \mathcal{T}_1$ is calculated as
\begin{equation}
        y_{\mathcal{T}_\iota+} = \frac{\sum\limits_{i\in \mathcal{T}_\iota} y_i w_i}
    {\sum\limits_{i \in \mathcal{T}_\iota}w_i},
\end{equation}
for $\iota =1,2$. Based on this, the between-subgroup heterogeneity is defined as
\begin{equation}
    Q_B(\mathcal{T}_1, \mathcal{T}_2) = \sum\limits_{\iota=1}^{2}\sum\limits_{i\in \mathcal{T}_\iota} w_i(y_{\mathcal{T}_\iota+}-y_{++})^2,
\end{equation}
where $y_{++}$ denotes the weighted mean for the union of both groups.
In the fixed effect model, we have $\tau^2 = 0$, implying that $w_i = 1 / v_i$. Consequently, the statistic $Q_B$ can be maximized with respect to $(\mathcal{T}_1, \mathcal{T}_2)$ independently of $\tau^2$. This simplifies tree construction and leads to a splitting criterion that is structurally analogous to the one used in the original CART framework. In contrast, the random effects setting requires re-estimation of the heterogeneity component for all terminal nodes after each split as every split modifies the global variance structure. For computational reasons, the DerSimonian--Laird estimator \autocite{dersimonian1986Metaanalysis} is used for $\hat{\tau}^2$. The procedure is embedded in a sequential tree-growing algorithm. Further implementation details are documented by \citeauthor*{li2020Multiple}\autocite{li2020Multiple} and in the corresponding package documentation\autocite{R_metacart}.

Growing a meta-CART requires several computational hyperparameters corresponding to the standard options in \texttt{rpart.control}, which are kept at their default values. Meta-CART uses a pruning parameter $c > 0$ that governs subtree selection via a $c \cdot \mathrm{SE}$ rule based on cross-validation error \autocite{dusseldorp2010Combining}. Data-driven recommendations for selecting $c$ have been proposed by \citeauthor*{li2019Flexible}\autocite{li2019Flexible} and are given by
\begin{equation}
\label{eq:c_metaCART}
c_{\text{Fixed effect}} =
\begin{cases}
1, & k < 80 \\
0.5, & k \ge 80
\end{cases}
\quad
c_{\text{Random effects}} =
\begin{cases}
1, & k < 120 \\
0.5, & k \ge 120 .
\end{cases}
\end{equation}
Larger values of $c$ lead to stronger pruning, whereas smaller values result in less aggressive pruning.
By default, REmrt, for which intra-node variability is additionally estimated in the form of $\tau^2$, can employ a \textit{lookahead} search.
Instead of evaluating only the immediate next split, the lookahead search jointly considers combinations of the next two splits, which theoretically allows the detection of strong IEs even when the corresponding MEs are absent.
This comes at increased computational cost and is often not feasible for small sample sizes (in our study particularly for $k = 13$ and $k = 23$).
We therefore did not apply the lookahead search in any of our analyses. However, for larger sample sizes it may be useful for fitting individual meta-CARTs.

Additionally, meta-CART inherits the standard \texttt{rpart} CART mechanism for handling missing values. Observations with missing values on the primary split variable are routed using surrogate splits\autocite{breiman1984classification}. 
Consequently, in principle, no studies with missing covariate values need to be excluded. However, the performance depends on the underlying missingness structure and a detailed investigation of the missingness aspect is beyond the scope of this paper.

Formally, a CART model with additional random effects is suitable when the data is assumed to follow the model
\begin{equation}
    \boldsymbol{y}_i 
    = \sum_{d=1}^D c_d \, \mathbf{1}\{\boldsymbol{x}_i \in R_d\} 
    + u_i + \epsilon_i,
    \label{eq:mod_CART}
\end{equation}
where $R_1,\ldots,R_D$ partition the covariate space into $D \in \mathbb{N}$ disjoint, axis-aligned rectangles, each associated with a constant $c_d \in \mathbb{R}$. Meta-CART fits this piecewise-constant model while allowing it to capture unexplained between-study variability. The terminal nodes of a meta-CART are suggested to be used directly to explain the heterogeneity of the meta-analysis. Simulation studies have shown that this approach performs well for data strictly following the model from Equation~\eqref{eq:mod_CART}\autocite{li2017MetaCART,li2019Flexible}. However, the resulting parameters do not align with the usually reported and analyzed linear coefficient structure of meta-regression models such as those in Equations~\eqref{eq:REMA} and~\eqref{eq:mod_IE}. We therefore propose selecting the MEs in the linear model for all variables appearing anywhere in the CART, as well as the IEs for variables that occur jointly along any path of the tree. This procedure preserves the linear structure of the classical random effects model while leveraging the tree’s ability to detect complex interaction patterns. The utility of CARTs for detecting IEs has been emphasized repeatedly in the literature \autocite{lissa2017MetaForest, dusseldorp2014Combinations}. Importantly, the strategy also respects the marginality principle. Consequently, the resulting procedure leads to a model in the form of Equation~\eqref{eq:mod_IE} rather than \eqref{eq:mod_CART}. Nevertheless, the three approaches should in principle be able to approximate the linear structures. 
Moreover, CARTs can be fitted even in settings with large $p/k$ ratios and have relatively high power to identify influential variables. This can be advantageous even though the tree itself is not the final model. In theory, covariates that have no influence at all should also not appear as splitting variables in the CART. Thus, meta-CART should prevent spurious findings. 

Meta-CARTs can be used as a pre-selection method for IEs but also interpreted as a one-step solution. IEs are selected when they appear along one branch of the tree. This could also include IEs including more than two variables but we only investigate two-way interactions.

Finally, we note that if a metric variable is split more than once and the threshold of the intermediate split lies between those of the outer splits, this may indicate the presence of a quadratic effect. This provides substantial potential for detecting second or higher order effects and is conceptually comparable to identifying IEs, consistent with the marginality principle for higher order terms\autocite{morris2023Marginality}. A detailed investigation of such effects lies beyond the scope of the present paper.

\subsubsection{Stabilized trees}
Using single trees as in meta-CART has several drawbacks, for example, its instability\autocite{hastie2009elements, lissa2017MetaForest}. A typical solution is to average over multiple trees. With a combination of bootstrapping and aggregating 
(bagging), this approach is implemented in the \emph{random forest} algorithm\autocite{breiman2001Random}. An adaptation for heterogeneous meta analytic models is available in the \emph{metaforest} framework\autocite{lissa2020Small}.

A fixed number of bootstrap samples are drawn, and a separate tree is fitted to each sample. By additionally randomizing the set of candidate predictors at each split, the ensemble reduces correlation between trees, and averaging their predictions substantially lowers variance compared to a single CART. This ensemble structure enables the computation of variable importance measures. Because each tree is built on perturbed data and predictor subsets, the algorithm can quantify the contribution of each variable to impurity reduction or predictive performance across the forest. Moreover, the bootstrap-based construction yields out-of-bag observations for each tree, which allow internal performance assessment and the computation of certain variable importance measures. 

Metaforest relies on these importance measures to assess the relevance of individual covariates, but does not provide results in the form of a classical meta-regression model as in Equation~\eqref{eq:REMA}, nor does it directly identify IEs as defined in Equation~\eqref{eq:mod_IE}. 
Adapting this approach for IEs is not straightforward. The random selection of variables at each split may separate interacting variables, making joint detection difficult. In addition, assessing the importance of an IE requires not only one correct split, but two splits occurring in the appropriate location within the tree. Therefore, we do not rely on the averaging aspect of classical random forests and instead fit meta-CARTs on $B \in \mathbb{N}$ bootstrap samples. Subsequently, the variables are selected following the idea of stability selection\autocite{meinshausen2010Stability}. Therefore, a threshold value $\lambda \in (0,1)$ is set and the matrix of relative selection frequencies is denoted by $\boldsymbol{A}$. The $i$th diagonal element $a_{ii}$ represents the relative frequency of trees in which the $i$th variable occurs, while the off-diagonal element $a_{ij}$ denotes the relative frequency of trees in which the $i$th and $j$th variable occur jointly along one path.
The $i$th ME is selected if $a_{ii} > \lambda$, and the IE of variable $i$ and $j$ when additionally
\begin{equation}
    \frac{a_{ij}}{\min(a_{ii}, a_{jj})} > \lambda.
\end{equation}
These rules can also be changed in accordance with the stability selection principle. In the following, we initially fix $\lambda = 0.5$ or analyze the frequencies separately. In our simulation study, the effect of $\lambda$ is explicitly investigated for random effects trees. 

Since $\lambda$ conceptually represents a global complexity parameter, we set the pruning parameter $c$ of the individual trees to $c = 0$ within the stabilized tree frameworks.
This corresponds to the $0 \cdot \mathrm{SE}$ rule, under which the tree with the smallest cross-validated error is selected from the pruning sequence, rather than disabling pruning entirely.
This choice is motivated by standard practice in random forests, where individual trees are grown with minimal constraints on their complexity, as overfitting is mitigated by averaging across many trees rather than by pruning each tree individually.
Further reducing pruning by controlling the underlying \texttt{rpart} hyperparameters (e.g., \texttt{cp}, \texttt{maxdepth}, \texttt{minsplit}, \texttt{minbucket}) directly only increased the Type~I error in a small side analysis.
Where this was accompanied by lower Type~II errors, it did not reflect an actual gain in selection quality but rather resembled a shift of $\lambda$.
In addition, it came with a considerable increase in running time for the ensembles.

Similarly, we do not employ the lookahead search for the individual REmrt trees within S-REmrt.
While lookahead is in principle applicable for larger sample sizes, its benefit is, to some extent, approximated by aggregating across the bootstrap samples.
Moreover, its computational cost would be substantial given the large number of trees fitted within the stabilized framework.

\section{Data-application - re-analysis of Kimmoun et al.}
\label{sec:re-analysis}
In this section we present an empirical application of the variable selection procedures introduced above. Using a large meta-analysis as a motivating example, we systematically assess how different methods perform.
\subsection{Motivating example and data description}

\citeauthor*{kimmoun2021Temporal}\autocite{kimmoun2021Temporal} conducted a large meta-analysis on acute heart failure 
which has been cited more than 120 times according to Google Scholar. While the authors reported, among other findings, a decrease in one-year mortality over time\autocite{kimmoun2021Temporal}, a subsequent re-analysis by \citeauthor*{knop2023Consequences}\autocite{knop2023Consequences} demonstrated that this apparent temporal trend may be confounded by variations in average patient age across studies. In the present example, we examine whether this conclusion would also have been supported by the variable selection procedures introduced above. Particular attention is paid to the corresponding IE, but also to the identification of potentially other IEs or MEs. A special focus is placed on tree-based selection methods.

The dataset is publicly available and contains a large set of study-level covariates as well as several potential endpoints (\url{https://osf.io/cxv5k/}). A descriptive summary of the dataset, including the number of missing values for all originally included studies, can be found in the Supplementary Materials. 
The outcome variables and possible metric moderators are listed in Table~S1, while potential binary moderators are presented in Table~S2.
Following the recent re-analysis by \citeauthor*{knop2023Consequences}\autocite{knop2023Consequences}, we focus on the logit-transformed one-year mortality rate as target variable, which is available for 204 observations. Some endpoints refer to the same underlying study but were measured in different years. In line with previous analyses, these observations are treated as independent studies. 
Among the 204 studies, seven substantively meaningful covariates exhibit either no missing values or only a moderate amount of missingness. The largest acceptable number of missing values occurs for the variable \textit{Discharge} (34 missing values). The eighth largest amount of missingness is observed for \textit{systolic blood pressure}, with 190 missing values. The categorical variable continent, comprising six levels, was excluded due to sparse category frequencies and the absence of a principled regrouping scheme in the absence of additional background information. The remaining six covariates are listed in Table~\ref{tab:covs}, together with short descriptions and the corresponding numbers of missing values, after restriction to the 204 studies with an available one-year mortality outcome.

Although univariate testing procedures and tree-based methods are generally capable of handling missing values in covariates, all analyses are conducted on a complete-case dataset to ensure comparability across methods. From a substantive perspective, the selection of covariates was guided by the assumption that all included variables may influence the outcome and that IEs may arise for several combinations of these covariates.

\begin{table}[htb]
    \caption{Included covariates from the \citeauthor*{kimmoun2021Temporal}\autocite{kimmoun2021Temporal} meta-analysis with their respective explanation, measurement level, value range, and number of missing values.} 
    \centering
    \begin{tabular}{lllcr}
        \toprule
        \textbf{Variable} & \textbf{Explanation} & \textbf{Measurement} & \textbf{Range of values} & \textbf{\#NAs}  \\ 
        \midrule
        \textit{Time} & Year of recruitment & metric & \{1981, ... , 2014\} & 0 \\
        \textit{Multi} & Multi- vs.\ monocentric & binary & multi/mono & 0 \\
        \textit{Trial} & Trial vs.\ survey study & binary & trial/survey & 0 \\
        \textit{Male (rate)} & Proportion of male patients & metric (rate) & \{0.075, ... , 0.986\} & 16 \\
        \textit{Age} & Mean age of patients & metric & \{51.86, ... , 89.00\} & 23 \\
        \textit{Discharge} & Outcome after discharge & binary & yes/no & 34 \\
        \bottomrule
    \end{tabular}
    \label{tab:covs}
\end{table}

We applied uni-variate testing (uni-test), multivariate testing (multi-test), the AICc-criterion, the BIC-criterion, fixed effect meta-CARTs in the singular (FEmrt) and stabilized version (S-FEmrt), and random effect meta-CARTs in the singular (REmrt) and stabilized version (S-REmrt). 
We concentrate on the one-stage variable selection procedures to avoid complications that may arise for multiple method approaches. Accordingly, the selected IEs can also be interpreted with respect to a pre-selection approach.
For the tuning parameters, we use the respective default settings without further optimization. These are $\alpha = 0.05$ for the testing procedures and $c$ as given in Equation~\eqref{eq:c_metaCART} for the single trees. For the stabilized tree methods, different values of the stabilization parameter $\lambda$ are discussed. The number of trees was set to $B = 1{,}000$. Smaller values, such as $B = 500$ or $B = 100$, yielded very similar results.
All metric variables are centered and standardized prior to analysis.

\subsection{Results of the re-analysis}

First, we emphasize that the choice between fixed effect and random effects should be guided by conceptual considerations. We opted for random effects in order to capture additional heterogeneity. As tree algorithms can vary substantially, we also included the so-called fixed effect meta-CARTs, as our goal is to investigate methodological differences. 
The selected parameters together with their estimates are reported for each method in Table~\ref{tab:method_comparison}. It should be noted that the parameter estimates are to be interpreted on the logit-transformed scale. Within the table, the S-FEmrt and S-REmrt results are evaluated for $\lambda = 0.5$. A detailed discussion of alternative $\lambda$ values is provided below. Visualizations of the fixed effect and random effects meta-CARTs are presented in Figures~S1 and~S2 of the Supplementary Materials. 

Table~\ref{tab:method_comparison} shows that the variable \textit{Age} is selected by every method. This is followed by \textit{Discharge} and the \textit{Male} rate, which are selected by all methods except both single-tree approaches, FEmrt and REmrt, resulting in six selections overall. The variable \textit{Time} is selected by only five methods, namely the testing approaches, AICc and both fixed effect tree versions. For the ME, these findings are in line with \citeauthor*{knop2023Consequences}\autocite{knop2023Consequences}, who also identify \textit{Age} as an influential variable.
However, the IE between time and age (\textit{Time:Age}) discussed in that study is selected by only three methods, including the S-FEmrt method, which by far selects the largest number of IEs. In total, eight IEs are selected at least once. The only additional IE selected three or more times is \textit{Age:Discharge}. Notably, this interaction is also selected by methods that, under the present settings, tend to select fewer IEs, as will become apparent in the subsequent simulation study.

\begin{table}[hb]
\caption{Estimates of the selected parameters in the re-analysis of \citeauthor*{kimmoun2021Temporal}\autocite{kimmoun2021Temporal} (S-FEmrt and S-REmrt with $\lambda = 0.5$)}
\label{tab:method_comparison}
\centering
\begin{tabular}[t]{llrrrrrrr}
\toprule
  & Uni-test & Multi-test & AICc & BIC & FEmrt & REmrt & S-FEmrt & S-REmrt\\
\midrule
(Intercept) & -1.00 & -0.90 & -0.91 & -0.99 & -1.13 & -1.11 & -1.01 & -1.00\\
\midrule
\textit{Time} & -0.04 & -0.07 & -0.06 &  & -0.04 &  & -0.10 & \\
\textit{Discharge} & -0.26 & -0.31 & -0.30 & -0.25 &  &  & -0.27 & -0.25\\
\textit{Age} & 0.18 & 0.05 & 0.09 & 0.23 & 0.23 & 0.21 & 0.26 & 0.24\\
\textit{Male} & -0.10 & -0.24 & -0.23 & -0.12 &  &  & -0.04 & -0.09\\
\textit{Multi} &  & -0.13 & -0.11 &  &  &  &  & \\
\textit{Trial} &  & -0.06 &  &  &  &  &  & \\
\midrule
\textit{Time:Age} & -0.08 &  &  &  & -0.09 &  & -0.12 & \\
\textit{Age:Multi} &  & 0.26 & 0.26 &  &  &  &  & \\
\textit{Male:Multi} &  & 0.22 & 0.21 &  &  &  &  & \\
\textit{Age:Disc.} &  &  & -0.06 & -0.12 &  &  & -0.08 & -0.14\\
\textit{Time:Disc.} &  &  &  &  &  &  & 0.08 & \\
\textit{Time:Male} &  &  &  &  &  &  & -0.12 & \\
\textit{Disc.:Male} &  &  &  &  &  &  & -0.05 & -0.05\\
\textit{Age:Male} &  &  &  &  &  &  & -0.04 & -0.02\\
\bottomrule
\end{tabular}
\end{table}

\begin{figure}[htbp]
  \centering
  \includegraphics[width=0.9\textwidth]{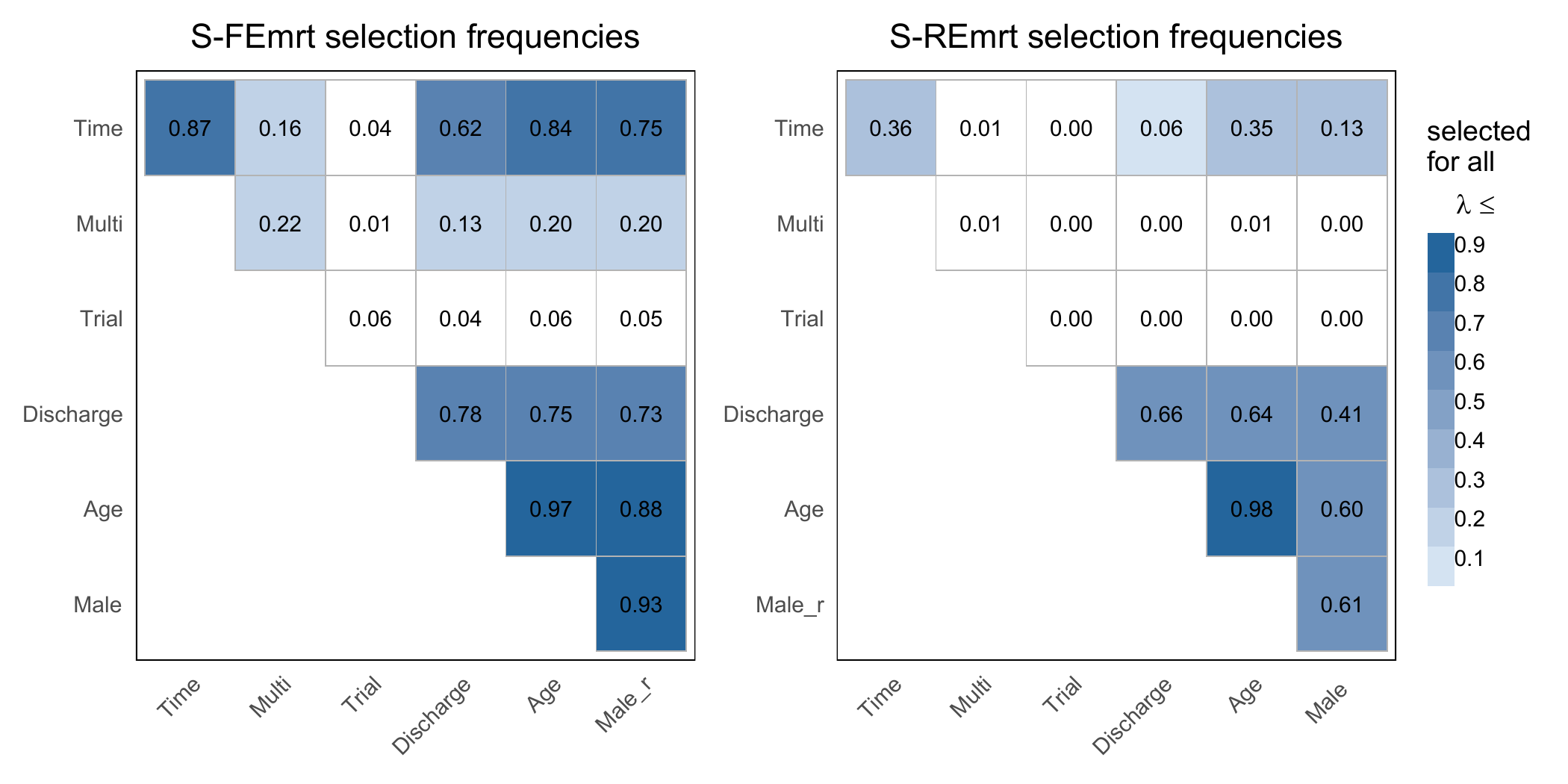}
  \caption{Selection frequency matrix $\boldsymbol{A}$ of the S-FEmrt and S-REmrt applied on the re-analyzed data set. The colors indicate the value of $\lambda$ at which each ME (main diagonal) or IE (off-diagonal) is selected}
  \label{fig:A_matrix}
\end{figure}

Instead of evaluating the stabilized meta-CARTs using a single cut-off value $\lambda$, one can instead examine the selection matrix $\boldsymbol{A}$. This matrix provides deeper insight into the underlying data structure. The matrices for S-FEmrt and S-REmrt in this re-analysis are shown in Figure~\ref{fig:A_matrix}. The numbers indicate the entries of the matrix $\boldsymbol{A}$, while the colors reflect the $\lambda$ value at which a ME or IE is selected. The tendency of S-FEmrt to select substantially more effects is further reinforced. Additionally, both matrices show that the selections are relatively clear-cut. All effects lie well above or well below the selected threshold of $\lambda = 0.5$. In both cases, \textit{Age} is the most frequently selected variable and also appears most often in interaction terms. This pattern also holds, albeit at a slightly lower level, for \textit{Discharge} and \textit{Male}. 
The largest difference between the two methods is observed for \textit{Time}. The ME of \textit{Time} is selected in 87\% of cases by S-FEmrt, compared to only 36\% by S-REmrt. This difference is also present for most IEs associated with \textit{Time}.

\section{Simulation study}

Several recent simulation studies address variable selection in meta-analytic models or analyze the performance or tree-based methods. However, these studies differ in scope and focus. For example, the study by \citeauthor*{cinar2021Using}\autocite{cinar2021Using} was not intended to incorporate tree-based methods or IEs. The studies by \citeauthor*{li2019Flexible}\autocite{li2017MetaCART, li2019Flexible} compare and fit models that do not necessarily assume a classical linear random effects structure. Notably, the simulation designs of \citeauthor*{li2019Flexible}\autocite{li2017MetaCART, li2019Flexible} rely on binary splits and tree-like structures, which naturally align with the structure assumed by meta-CART-type methods. In the study of \citeauthor*{lissa2017MetaForest}\autocite{lissa2017MetaForest}, which combines moderator selection based on feature importance with a focus on predictive performance evaluated via cross-validated $R^2$, the identified moderators are not interpreted as parametric IEs in the sense of Equation~\eqref{eq:mod_IE}.
Thus, a study based on the classical random effects meta-regression model, explicitly incorporating parametric and interpretable IEs, is needed. The simulation study was performed in \texttt{R} (version 4.4.2)\autocite{R} with special use of the packages \texttt{metacart} \autocite{R_metacart} and \texttt{metafor}\autocite{viechtbauer2010Conducting}.

Linear methods considered are univariate and multivariate testing, both with $\alpha = 0.05$ as well as the AICc and BIC criterion. The single-tree methods comprise fixed effect and random effects meta-CARTs (FEmrt and REmrt) with pruning rules following Equation~\eqref{eq:c_metaCART}. Stabilized tree methods are stabilized fixed effect and random effects meta-CARTs (S-FEmrt and S-REmrt), both initially applied with $\lambda = 0.5$. A subsequent analysis of different $\lambda$-values within the S-REmrt is given in the last part of the simulation study (Section~\ref{sec:sim_lambda}). The number of trees is set to $B = 100$. As the re-analysis presented in Section~\ref{sec:re-analysis} showed very similar results for $B=100$ and $B = 1000$, this choice substantially reduces computational cost without a noticeable loss in performance.

The analysis of the simulation results is divided into three steps. First, we evaluate how well the methods perform, when the DGM is strictly linear. In a second step, we examine how the results differ when there are small deviations from this structure. We explicitly disregard more complex tree-style DGMs, as such settings have been studied in the context of tree-based methods and, moreover, the classical linear random effects model would not be appropriate in this case. In addition, a practical meta-analysis under such conditions, especially with a limited amount of data, would not be expected to yield reliable results. Finally we discuss the influence of the selection threshold $\lambda$ within the S-REmrt.

\subsection{Simulation design}

To generate comparable and neutral settings, we follow the general approach of \citeauthor*{cinar2021Using}\autocite{cinar2021Using} and base our study on an empirical dataset. Specifically, we use the \citeauthor*{kimmoun2021Temporal}\autocite{kimmoun2021Temporal} dataset from the previous section as it contains many potential covariates and a relatively large number of studies to select from. To retain a realistic joint distribution of covariates, we do not generate them synthetically. Instead, we sample directly from the available studies (with replacement), which is valid given the relatively large number of studies representing an appropriate distribution. The outcome variable is then generated synthetically using controllable coefficients. This procedure corresponds to a \textit{statistical plasmode simulation} with synthetic outcome generation\autocite{schreck2024Statistical}.

The full dataset consists of 335 studies. Missing covariate values are imputed by randomly sampling observed values from the same variable (hot deck imputation). 
This is performed only for variables with a sufficient number of original observations.
These include all variables listed in Table~\ref{tab:covs} as well as the metric variable \textit{systolic blood pressure} (SBP), which originally has 190 missing values. 
Since this preserves the marginal distributions and draws from a relatively large number of observations, it is not expected to substantially distort the correlation structure. 
The resulting correlation structure (after imputation) is shown in Table~S3 in the Supplementary Materials. All metric variables are centered and standardized prior to analysis.

We simulate the target variable logit-transformed one-year mortality rate using a controlled DGM. Specifically, the target variable is generated on $\mathbb{R}$ and linked to a set of covariates through regression models that map the linear predictor to combinations of MEs and IEs. Across simulation scenarios, we consider different functional forms for this mapping, including linear additive and non-linear structures.

Additionally, we conducted a classical simulation study in which covariates are generated in a controlled manner from a multivariate normal distribution, with standardized mean differences (SMDs) as the outcome.
This design choice is closer to other simulation studies in the meta-analytic literature \autocite{viechtbauer2015Comparison, lissa2017MetaForest, welz2020Simulation}.
In doing so, we lose the data structure of our motivating example but also complement the plasmode study by allowing us to vary the covariate distribution, particularly the balancedness of binary moderators and the correlation between moderators.
Results for these two factors are described in the following section.
Further details on the design and output materials can be found in Section~C of the Supplementary Materials.
Overall, the findings largely coincide with those of the main plasmode simulation study. \\

\noindent\textbf{Linear DGMs.} We first consider the strictly linear model from Equation~\eqref{eq:mod_IE}. Here, we define 14 different simulation settings, which differ in the way the parameter coefficients are chosen (see Table~\ref{tab:para_settings}). The magnitudes of the parameters are unified to prevent biased results for different IEs. The signs are aligned with the re-analysis in the previous section. We consider three types of IEs with meaningful interpretation: metric–metric (\textit{Time:Age}), binary–metric (\textit{Disc.:SBP}), and binary–binary (\textit{Multi:Trial}). Among the 14 settings, we define five different groups: one group contains no IEs (Settings~1 and~2), three groups contain exactly one of the three IEs (Settings~3--5, 6--8, and 9--11 one type of IE each), and one group includes all of the three IEs (Settings~12--14). For the group with no IEs, one setting is specified completely empty (Setting~1) and one contains all possible MEs (Setting~2). For each further group, we define a setting containing only the specific IE (Settings~3, 6, 9, and~12), one setting with additionally the corresponding MEs (Settings~5, 7, 10, and~13), and a setting with all additional MEs (Settings~5, 8, 11, and~14).

\begin{table}[hb]
\small
\caption{Fixed parameter coefficients for the different simulation settings. Empty grids or not included variables mean that no influence is modeled.}
\label{tab:para_settings}
\centering
\begin{tabular}{lccccccccccc}
\toprule
\textit{} & \multicolumn{8}{c}{\textit{ME}} & \multicolumn{3}{c}{\textit{IE}} \\
\cmidrule(lr){2-9} \cmidrule(lr){10-12}
 & \textit{Intercept} & \textit{Time} & \textit{Trial} & \textit{Male} & \textit{Age} & \textit{SBP} & \textit{Multi} & \textit{Disc.} & \textit{Age:Time} & \textit{Disc.:Time} & \textit{Disc.:Multi}\\
\midrule
1 & -1 &  &  &  &  &  &  &  &  &  & \\
2 & -1 & -0.5 & -0.5 & -0.5 & 0.5 & 0.5 & -0.5 & -0.5 &  &  & \\
\hdashline[1pt/2pt]

3 & -1 &  &  &  &  &  &  &  & -0.5 &  & \\
4 & -1 & -0.5 &  &  & 0.5 &  &  &  & -0.5 &  & \\
5 & -1 & -0.5 & -0.5 & -0.5 & 0.5 & 0.5 & -0.5 & -0.5 & -0.5 &  & \\
\hdashline[1pt/2pt]

6 & -1 &  &  &  &  &  &  &  &  & -0.5 & \\
7 & -1 & -0.5 &  &  &  &  &  & -0.5 &  & -0.5 & \\
8 & -1 & -0.5 & -0.5 & -0.5 & 0.5 & 0.5 & -0.5 & -0.5 &  & -0.5 & \\
\hdashline[1pt/2pt]

9 & -1 &  &  &  &  &  &  &  &  &  & 0.5\\
10 & -1 &  & -0.5 &  &  &  & -0.5 &  &  &  & 0.5\\
11 & -1 & -0.5 & -0.5 & -0.5 & 0.5 & 0.5 & -0.5 & -0.5 &  &  & 0.5\\
\hdashline[1pt/2pt]

12 & -1 &  &  &  &  &  &  &  & -0.5 & -0.5 & 0.5\\
13 & -1 & -0.5 & -0.5 &  & 0.5 &  & -0.5 & -0.5 & -0.5 & -0.5 & 0.5\\
14 & -1 & -0.5 & -0.5 & -0.5 & 0.5 & 0.5 & -0.5 & -0.5 & -0.5 & -0.5 & 0.5\\
\bottomrule
\end{tabular}
\end{table}

To incorporate heterogeneity, we simulate five levels: no heterogeneity ($\tau^2 = 0$) and the confidence interval bounds for $\tau^2$ from the model in Section~\ref{sec:re-analysis}, estimated once including all available covariates ($\tau^2 = 0.141, 0.233$), and once excluding all covariates ($\tau^2 = 0.195, 0.317$). Based on the study by \citeauthor*{geissbuhler2021Most}\autocite{geissbuhler2021Most}, which examines sample sizes in meta-regressions, we subsample $k = 13$, $k = 23$, and $k = 41$ studies, corresponding to the 0.25 quantile, the median, and the 0.75 quantile reported therein. Moreover, we additionally consider a larger subsample of $k = 100$ studies.
Each of the 14 parameter settings is combined with all four values of $k$ and all five values of $\tau^2$, resulting in a total of 280 settings. Each setting is replicated 100 times to balance Monte-Carlo precision and computational cost. 

No additional factors are varied, as initial screening analyses showed only predictable effects consistent with general findings in variable selection and meta-analysis, such as improved performance with larger study sizes or fewer potential covariates. 

The sampling variance $v_i$ is generated using the large-sample approximation under a binomial model with the assumed logit link. 
The target variable is transformed via the inverse logit to obtain $p_{\text{true}}$, and the variance is then computed as
\begin{align}
    v_i = \frac{1}{n_i \, p_{\text{true}} (1 - p_{\text{true}})} ,
\end{align}
making it directly dependent on both study size and the underlying event probability. 
The study-specific sample size $n_i$ is taken from the bootstrapped rows of the original dataset, preserving the empirical distribution of sample sizes. 
Up to now, we focused on a classical linear additive underlying model. As a sensitivity analysis to this assumption, we additionally define two setups in which the IE is non-linear. \\

\noindent\textbf{Non-Linear DGMs.} In these scenarios, the data is no longer generated from the model in Equation~\eqref{eq:mod_IE}, but instead follows an interaction structure that is not globally linear, but can be approximated by recursive partitioning as formalized in Equation~\eqref{eq:mod_CART}. The parameter specifications are taken from Settings~3--8 in Table~\ref{tab:para_settings}. The only difference concerns the specification of the IEs. Specifically, the \textit{Time:Age} effect is now defined as
\begin{equation}
\beta \cdot x_{\text{Time}} \cdot x_{\text{Age}} \cdot I(x_{\text{Age}} > \overline{x}_{\text{Age}}),    
\end{equation}
and the \textit{Disc.:Time} effect is defined as
\begin{equation}
\beta \cdot x_{\text{Disc.}} \cdot I(x_{\text{Time}} > \overline{x}_{\text{Time}}).    
\end{equation}
Combined with the ME specifications from Settings~3--8, this leads to six non-linear settings. Each one is again combined with all $k$ and $\tau^2$ values resulting in 120 additional simulation settings.

\subsection{Simulation results}
We visualize and discuss both Type~I and the Type~II error of IE selection. Specifically, we evaluate errors with respect to the specification of $\boldsymbol{L}$ defined for Equation~\eqref{eq:mod_IE}. This does not evaluate whether the variable is significant in the final model, as valid inference for such variables is a separate topic itself. Issues regarding overoptimistic results are briefly discussed in Section~\ref{sec:discussion}.

\subsubsection{Detecting IEs for linear DGMs - Type~I and II errors}

\begin{figure}[htbp]
  \centering
  \includegraphics[width=0.9\linewidth]{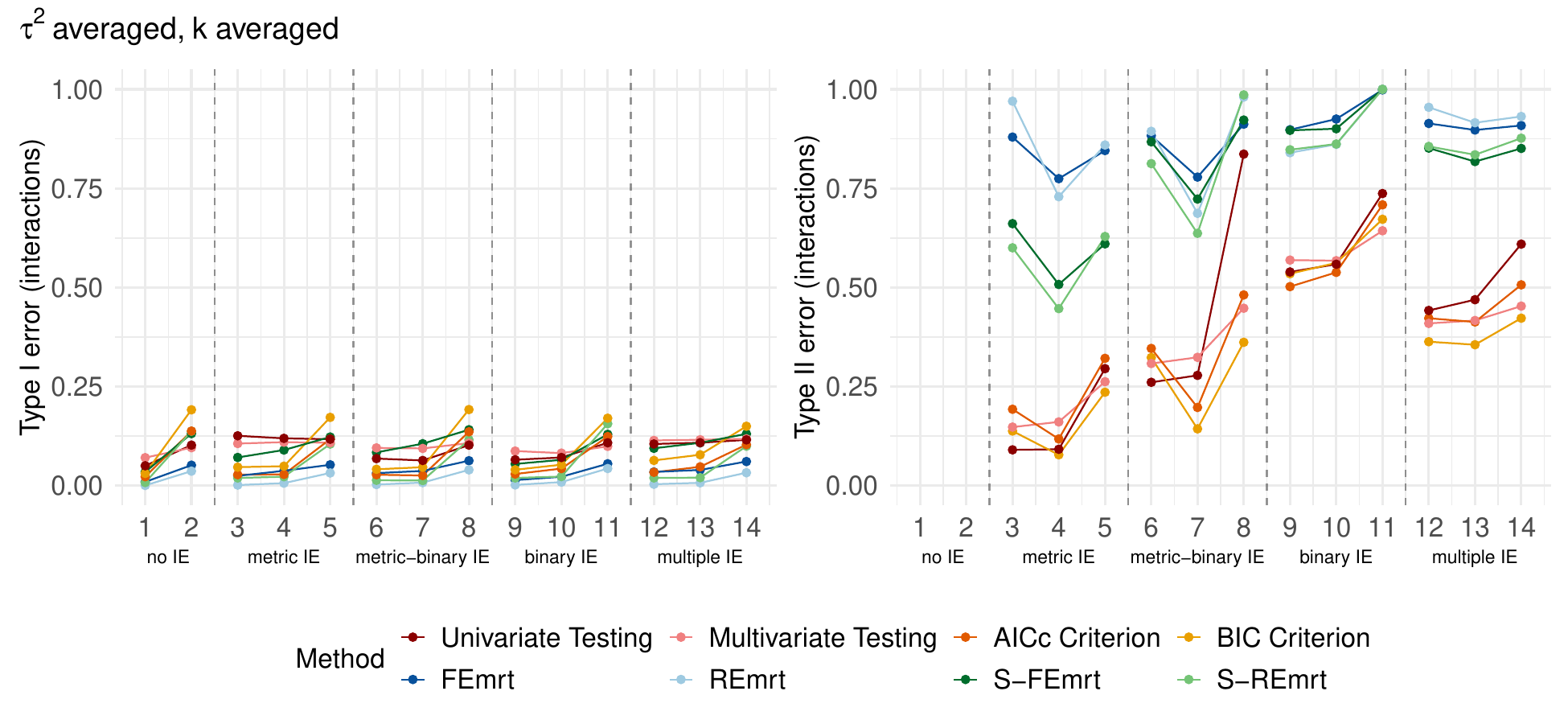}
  \caption{Average Type~I (left) and Type~II (right) IE selection error averaged across all settings for $k$ and $\tau^2$, visualized for all 14 linear parameter set-ups}
  \label{fig:combined_plot_averaged}
\end{figure}

Type~I and Type~II errors, averaged across all values of $\tau^2$ and $k$, are shown in Figure~\ref{fig:combined_plot_averaged} for all linear Settings~1--14. The results for the most extreme values of $k$ and $\tau^2$ ($k= 13$ and $k = 100$ as well as $\tau^2 = 0$ and $\tau^2 = 0.317$) are shown separately in Figure~\ref{fig:combined_plot_extreme}.
Detailed plots for all values of $\tau^2$ and $k$ are provided in the Supplementary Materials (Figures~S4 to S23) and exact numbers are listed in Tables~S5 to S12. 
No Type~II errors are reported or plotted for Settings~1 and~2, as no IEs are present in these scenarios.\\

\noindent
\textbf{Averaged Results.} Since the results mostly have some consistent trends for varying $\tau^2$ and $k$, and a full factorial design is available, we first discuss errors averaged across all settings. Here, all methods control the Type~I error considerably well (left panel of Figure~\ref{fig:combined_plot_averaged}).

Among the linear selection methods, performance regarding both error types is similar in most settings.
Notably, univariate testing performs worse regarding Type~II error for settings with many influential predictor variables (especially Setting~8) and many IEs (Settings~12--14), while also showing a relatively high Type~I error rate together with multivariate testing.
The criterion-based methods tend to achieve a better balance between Type~I and Type~II error.
Here, AICc is somewhat more conservative and BIC somewhat more liberal, which often leads to the lowest Type~II error rate for the BIC in Settings~5--8 and 12--14, but also, especially for settings with many variables, to an increased Type~I error rate (Settings~2, 5, 8, 11, 14).

Tree-based approaches exhibit substantially larger Type~II error rates across most settings. Methods based on stabilized trees improve performance especially for metric IEs, whereas only moderate improvements are observed for binary and multiple IEs. The Type~I error is also low for all tree-based methods, and is in most settings lower for the S-REmrt than for the S-FEmrt.
Thus, among the tree-based methods, the S-REmrt approach achieves the lowest Type~II error rates while also reducing Type~I errors relative to the S-FEmrt, making it the preferable choice.

\begin{figure}[htbp]
  \centering
  \includegraphics[width=0.8 \linewidth]{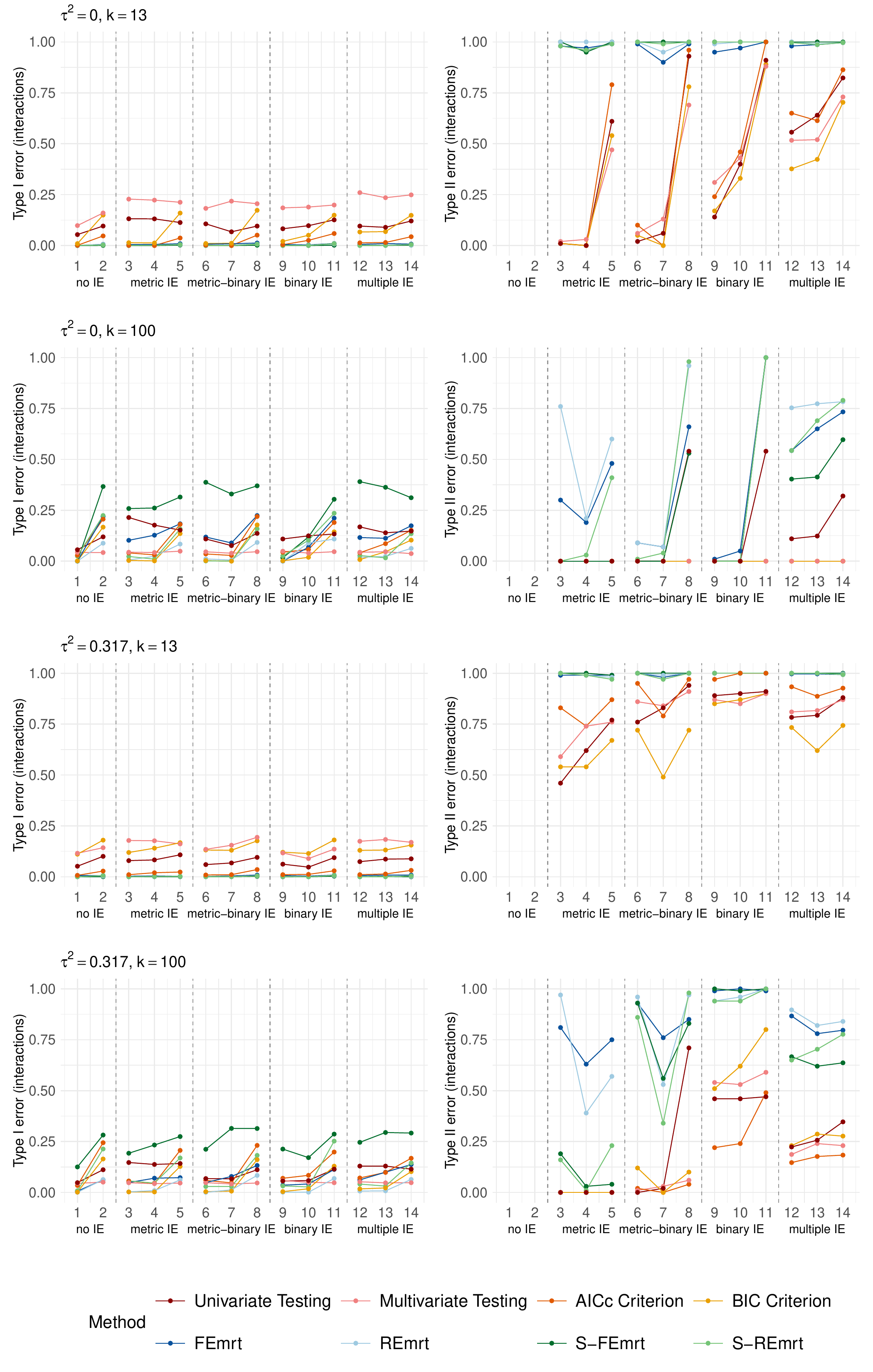}
  \caption{Type~I (left) and Type~II (right) IE selection errors for all combinations of the most extreme $\tau^2$ and $k$ values and all 14 linear set-ups; $k = 13,100$, $\tau^2 = 0,0.317$}
  \label{fig:combined_plot_extreme}
\end{figure}

\noindent\textbf{Impact of $k$.}
Separating the averaged results by the number of studies $k$ provides a clear explanation for the performance of the tree-based approaches.
When $k$ is small, almost no IEs are detected at all.
They start to detect interactions in our settings from $k = 23$ onwards, especially for low heterogeneity.
This complements the recommendations of \citeauthor*{li2017MetaCART}\autocite{li2017MetaCART}, who suggest that meta-CART benefits from meta-analyses with more than $40$ studies.
As for larger $k$ values, Type~II error rates decrease, particularly for both stabilized tree versions, while at the same time the Type~I error rate strongly increases for the S-FEmrt.
The best balance here is again found by the S-REmrt.

The classical linear procedures also benefit from larger $k$, but achieve overall better results for smaller values of $k$.
Notably, for $k = 13$, at comparable Type~II error rates, univariate testing has lower Type~I error rates than multivariate testing, whereas the reverse holds for $k = 100$.

\noindent\textbf{Impact of $\tau^2$.} The impact of $\tau^2$ again has to be analyzed separately for tree-based and linear approaches.

For the tree-based methods, the impact of $\tau^2$ is small when the number of studies is low.
In this case, few to almost no IEs are selected for all $\tau^2$ levels in both cases.
The differences become more pronounced for large values of $k$.
Thus, comparing the plots in the second and fourth row of Figure~\ref{fig:combined_plot_extreme}, we observe only minor changes of the Type~I error rates with $\tau^2$, whereas Type~II error rates are slightly lower, and for binary IE even largely lower, for a low level of heterogeneity.

For the group of linear methods, Type~I errors remain fairly similar for large and low levels of $\tau^2$.
Regarding Type~II error, a clear deterioration can be observed.
For $k = 100$ and $\tau^2 = 0$, all linear methods except univariate testing correctly detect all IEs (see second row). 
This is no longer the case for $\tau^2 = 0.317$.
For $k = 13$, almost no IEs are detected anymore for the binary IEs, in particular for AICc.
Related work has studied variable selection with classical linear procedures for different settings of $k$ and $\tau^2$ \autocite{cinar2021Using}, as well as more general inference approaches in the presence of increasing heterogeneity, including robust testing alternatives \autocite{thurow2024Robust, welz2020Simulation, viechtbauer2015Comparison}.

\subsubsection{Detecting IEs in non-linear DGMs - Type~I and II errors}
Type~I and Type~II errors for the six settings with the non-linear IEs are averaged across all $\tau^2$ and $k$ values and visualized in Figure~\ref{fig:combined_plot_non_linear}. 
Non-averaged plots and exact numbers for each method can be found in the Supplementary Materials (Figures~S24 to S42 and Tables~S13 to S20).

\begin{figure}[htbp]
  \centering
   \includegraphics[width=0.8  \linewidth]{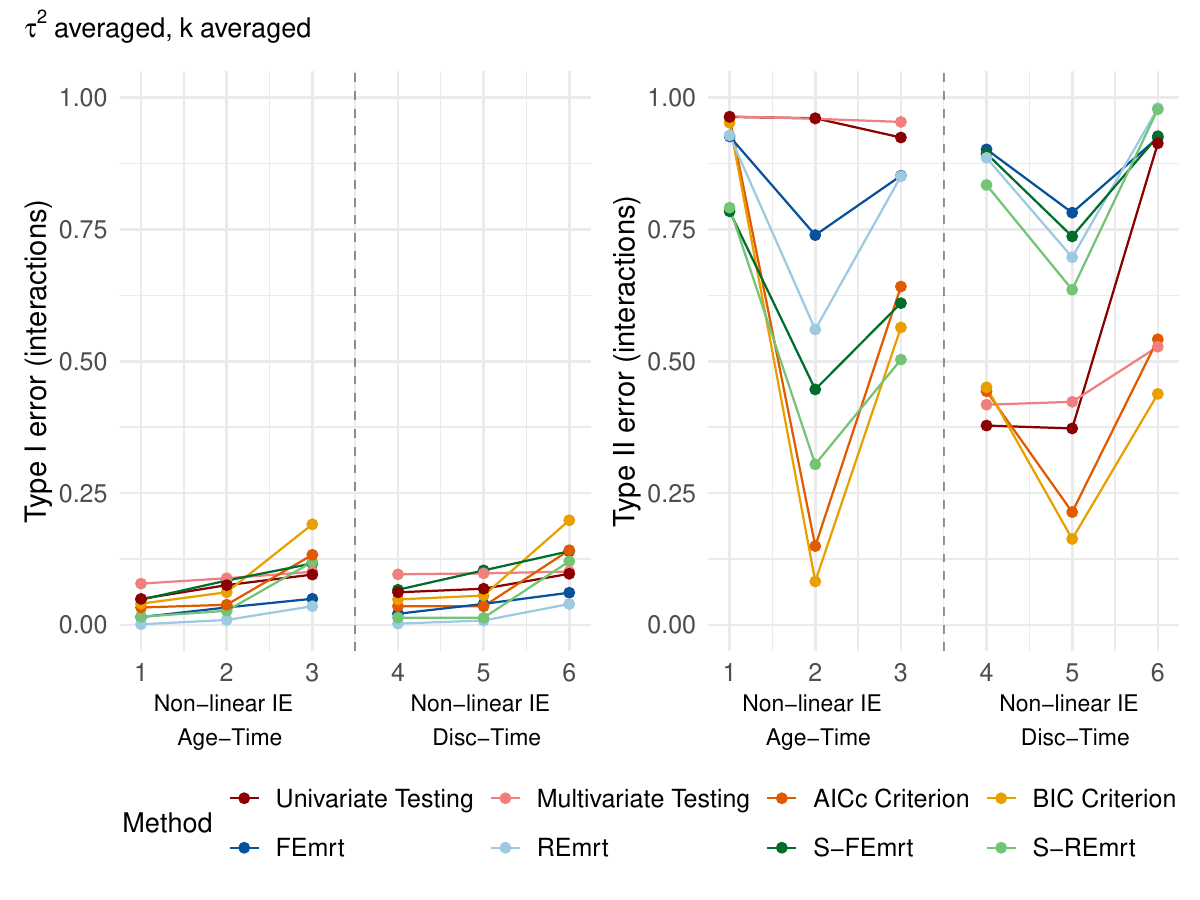}
  \caption{Average Type~I (left) and Type~II (right) IE selection error across all settings for $k$ and $\tau^2$ for all six non-linear parameter set-ups}
  \label{fig:combined_plot_non_linear}
\end{figure}

Considering first the linear methods, they show a loss in performance compared to the respective settings in the linear case (settings~3--5 and 6--8 in Figure~\ref{fig:combined_plot_averaged}), especially with respect to Type~II error. 
For the criteria approaches, this loss is only minimal while it is substantial for the test-based approaches.
A possible explanation is the difference in the underlying mechanisms, namely the likelihood-based comparison used by the criteria versus the Wald-type tests.

The performance of the trees remains at a similar level, meaning they are notably better than the test-based methods in settings~1--3. The S-REmrt even outperforms the criterion-based methods regarding Type~II error in settings~1 and~3.
For this type of metric IE, they therefore represent a useful alternative.
The results for the individual $\tau^2$ and $k$ values in the Supplementary Materials again show that the trees can perform better for individual settings with larger $k$. 
The averaged results shown in Figure~\ref{fig:combined_plot_non_linear}, however, also include scenarios with $k = 13$, in which the trees select almost no IEs. This inflates their averaged Type~II error rates performance 
and therefore underestimates their good performance in settings with larger $k$.

The difference in the trees' performance between the two non-linear IEs can be attributed to the splitting signal. For the second, \textit{Disc.-Time} IE, only a weak splitting signal is present, in contrast to the \textit{Age:Time} interaction present in the other non-linear IE.
This makes the former difficult to detect for greedy tree-growing procedures.
In such situations a lookahead strategy could potentially be useful, provided that the sample size is sufficiently large.

\subsubsection{Impact of \texorpdfstring{$\lambda$}{lambda} within the S-REmrt}
\label{sec:sim_lambda}

\begin{figure}[htbp]
  \centering
  \includegraphics[width=0.9\textwidth]{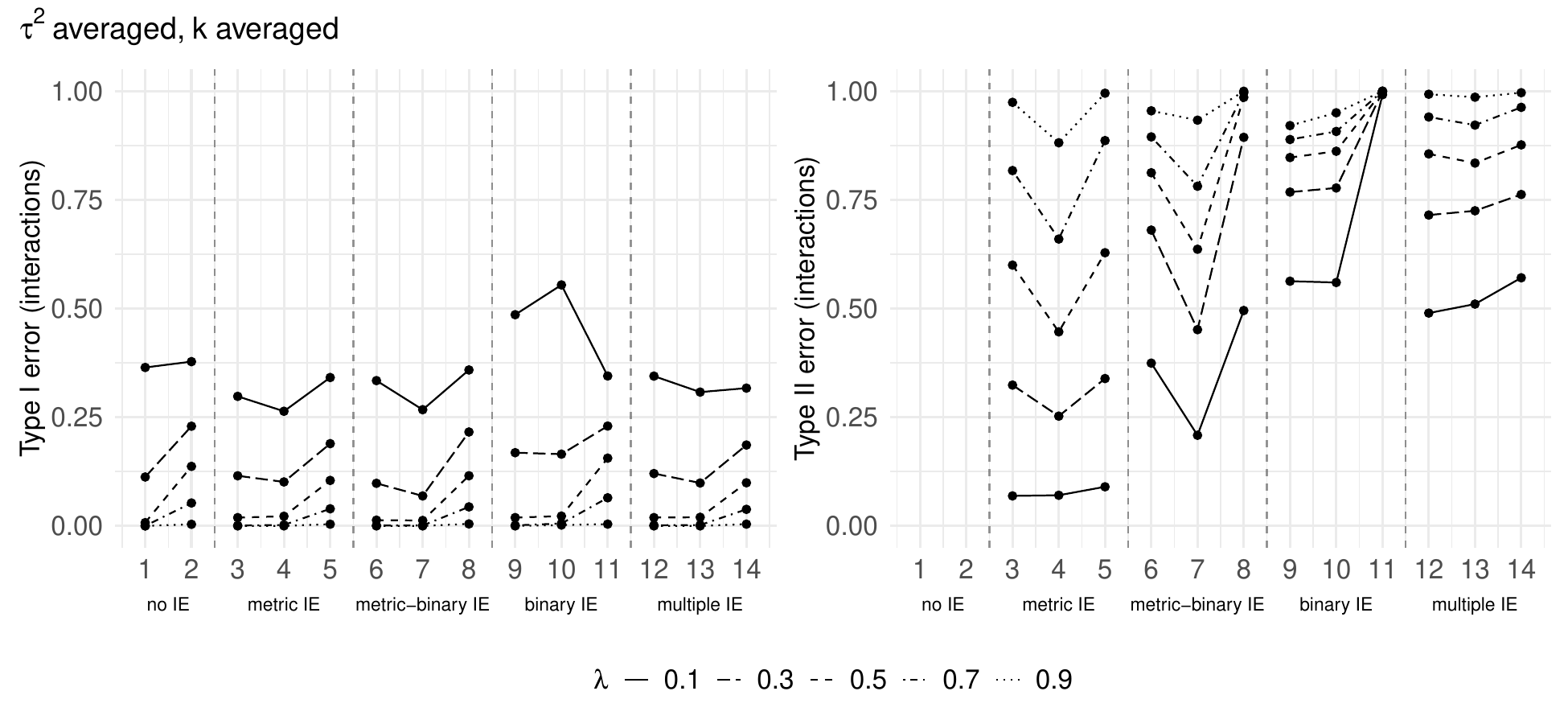}
  \caption{Average Type~I (left) and Type~II (right) IE selection error for the S-REmrt and different $\lambda$ values averaged across all settings for $k$ and $\tau^2$, visualized for all 14 linear parameter set-ups}
  \label{fig:combined_plot_averaged_lambda}
\end{figure}

The selection results depend on the choice of tuning parameters. Since tuning procedures for linear methods are well investigated and S-FEmrt has emerged as the inferior tree-based approach in many settings, we focus on a more detailed analysis of the influence of the parameter $\lambda$ within S-REmrt. Therefore, the results from the linear setting are re-evaluated for $\lambda = 0.1, 0.3, \ldots, 0.9$. The averaged results are shown in Figure~\ref{fig:combined_plot_averaged_lambda}. Exact values and results for individual combinations of $k$ and $\tau^2$ are provided in the Supplementary Materials (Figures~S44 to S62 and Tables~S21 to S25). 

The results indicate that the effect of $\lambda$ behaves as intended. Smaller values of $\lambda$ select a larger number of IEs and lead to higher Type~I error rates and lower Type~II error rates. Conversely, larger values of $\lambda$ result in lower Type~I errors and higher Type~II errors. Overall, very large values of $\lambda$ such as $\lambda = 0.9$ detect hardly any IEs, while very small values such as $\lambda = 0.1$ yield only marginal gains in Type~II error at the cost of a comparatively large increase in Type~I error. This trade-off is particularly pronounced for binary IEs in Settings~9--11. Values of $\lambda$ in the range $(0.3, 0.7)$ appear to provide a reasonable balance. In the previous simulations, we have therefore chosen $\lambda = 0.5$. For this value, S-REmrt attains a maximum averaged Type~I error of approximately $10\%$, which is comparable to the linear methods, particularly to the testing approaches conducted at the $\alpha = 0.05$ significance level, whose maximum averaged proportion of spurious findings is also around $10\%$. 

The choice between larger and smaller values of $\lambda$ should ultimately depend on the application context. In a pre-screening setting, smaller values of $\lambda$ may be preferable, whereas larger values of $\lambda$ can be advantageous when spurious findings are more crucial. In this context, it is also informative to consider the number of studies $k$. For small sample sizes, choosing a smaller value of $\lambda$ can lead to a substantially higher number of correctly identified IEs and may counteract the conservative nature of tree-based methods. In our simulation runs, good performance is observed from setting $k = 23$ onwards, particularly for smaller values of~$\lambda$.

With respect to the level of heterogeneity, no pronounced trends can be observed. In particular, the Type~I error rates remain very similar across all values of $\tau^2$. Only the Type~II error rates differ slightly. They are smaller when $\tau^2 = 0$, whereas for the remaining $\tau^2$ values they stay relatively constant.

For completeness, the corresponding results for the S-FEmrt are reported in the Supplementary Materials (Figures~S63 to S82 and Tables~S26 to S30). 
The effect of $\lambda$ is very similar. \\

\noindent\textbf{Recommended Choice of $\lambda$.} We deliberately refrain from numerically optimizing $\lambda$ or recommending a single universally optimal value. Type~I and Type~II errors are not directly comparable, rendering any optimization criterion ambiguous. Moreover, the behavior of these error rates is too heterogeneous across different settings. Our recommendation is therefore to either choose a central value such as $\lambda = 0.5$, or to explore multiple values of $\lambda$ in exploratory analyses. Alternatively, the selection matrix $\boldsymbol{A}$ (shown in Figure~\ref{fig:A_matrix}) can be inspected, as it allows characteristic patterns to be identified particularly clearly.

\subsubsection{Non-plasmode simulation study: summary of results}

The results from the fully synthetic simulation study are largely consistent with those from the plasmode study. A key extension is that the covariate distribution is controlled via additional parameters $\delta$ and $\rho$, where $\delta$ controls the balancedness of binary moderator variables and $\rho$ controls the covariance between moderators.
All results and the explicit simulation design description can be found in Section~C of the Supplementary Materials (linear Figures S83 to S87, non-linear Figures S88 to S92, lambda evaluation Figures S93 and S94).
 
Overall, we only found one notable difference: largely independent of $\rho$ and $\tau^2$, S-FEmrt tends to be slightly more liberal in some settings. This is often reflected in larger Type~I errors and, in some of the non-linear settings, also in lower Type~II errors.
Regarding $\delta$ and $\rho$, no remarkable differences can be detected.
One exception, as expected, is that the univariate testing procedure cannot handle correlation between moderators (at $\rho = 0.5$) well, especially for metric IEs.
However, all other methods, which are designed to handle moderators jointly, do not show notable differences.
Further aspects of the moderator distributions are outside the scope of this simulation.
However, aspects like slightly skewed or non-normal distributions were already covered in the plasmode simulation due to sampling from the empirical distribution.

\section{Discussion}
\label{sec:discussion}
In this paper, we analyzed how interaction effects (IEs) can be detected in meta-regressions. In many situations, especially when available studies are scarce, the general approach is to model only individual IEs that are justified by external domain knowledge. However, when too many IEs appear plausible or when an exploratory approach is taken, classical variable selection procedures reach their limits. This problem is particularly severe in meta-analyses due to the typically small number of studies and the presence of heterogeneity.

When dealing with IEs, an important modeling decision is whether the marginality principle is respected or not.\autocite{nelder1977Reformulation} Since 
interpretability is a central aspect in meta-analyses, 
we restrict our attention to the case where the marginality principle is enforced.
This strong focus on interpretability is also a reason why it is difficult to apply complex machine learning methods to meta-analyses. 
As a consequence, interpretable linear models remain the dominant modeling framework in meta-analysis.

At the same time, there are recommendations that tree-based methods can be helpful for the detection of IEs\autocite{lissa2017MetaForest, dusseldorp2014Combinations}. 
This motivated the question of how such approaches can be combined with the target structure of a linear model and how effectively such combinations, as well as classical linear variable selection methods, perform in this case.

Overall, we described and analyzed methods and strategies for handling (mixed effects) meta-regressions with many potential IEs while respecting the marginality constraint. Among procedures based on statistical tests, we considered univariate and multivariate testing, based on Wald-type tests.
Accordingly, multivariate testing was restricted to forward selection due to the requirement of handling a large number of possible parameters induced by the IEs. In addition, we investigated criterion-based methods, specifically the small-sample corrected AIC (\textit{AICc}) and the BIC, both of which were also implemented using a forward strategy. The tree-based methods were based on the \textit{meta-CART} framework\autocite{li2020Multiple} and included a random effects tree (REmrt) and a fixed effect version (FEmrt). Effects were interpreted as IEs when variables occurred along the same branch. Our analysis was restricted to two-way IEs. Since trees are known to be highly unstable, they are often combined into ensembles, such as random forests or, in the context of meta-analyses, approaches like the \textit{metaforest}\autocite{lissa2020Small}. Based on the ideas of stability selection\autocite{meinshausen2010Stability}, we therefore constructed ensembles of meta-CARTs using bootstrapped data, resulting in the S-REmrt and S-FEmrt approaches.

All methods were applied to a real data set based on the analysis of \citeauthor*{kimmoun2021Temporal}\autocite{kimmoun2021Temporal}, where a re-analysis of \citeauthor*{knop2023Consequences}\autocite{knop2023Consequences} indicated that confounding due to IEs may be present. In addition, we conducted a plasmode simulation based on this data set. We thereby investigated different data-generating models (DGMs): a classical linear model as well as two different models with non-linear IEs.

We overall found that linear methods perform best in situations with strictly linear IEs.
Tree-based methods become competitive only when the number of studies $k$ is sufficiently large. 
In our settings, this was observed from $k = 23$ onward and more consistently at $k = 41$. However, this is context dependent and also depends on the number of potential covariates and the distribution of the data. 
Furthermore, tree-based methods are a robust alternative when the IEs deviate from strict linearity.
Criterion-based selection, however, is also very effective, and even more effective in situations with relatively simple nonlinearities.
In the cases where tree-based methods are comparatively strong, this was primarily the case for the stabilized approaches, with S-REmrt showing superior performance. 
We therefore examined the impact of the regularization parameter $\lambda$ on S-REmrt in more detail in the simulation study.
Based on the simulation results, practical guidance is provided in the following and final Chapter~\ref{Sec:guidance}.

Throughout the paper, we discussed strategies for dealing with a large number of potential parameters, in particular those induced by IEs. This discussion included a two-step approach with a pre-selection step, as well as one-step modeling approaches that directly rely on the methods considered in this work. If a two-step approach with prior IE selection is applied instead of a one-step procedure, we highlight that incorporating tree-based methods may be particularly advantageous. At the same time, we caution that while two-stage procedures can reduce the parameter space and thereby enable a potentially better second procedure, they also entail a risk of over optimization.
This risk, namely that the nominal error level may not be maintained in the secondary procedure, arises both for classical selection methods and for tree-based selection methods.
For tree-based methods, we want to particularly highlight the case where the selected variables are themselves only constructed through the greedy search of a tree, for example, when testing the effect of a variable that has been dichotomized by the tree. \autocite{neufeld2022tree}
As originally proposed, meta-CART uses the terminal nodes themselves as new variables, which constitutes exactly such a case.
We therefore restrict our evaluation to whether the original, pre-specified variable, occurs anywhere in the tree, rather than modeling tree-derived dichotomizations.

Regarding classical linear selection, we want to stress that the differences between testing and criterion methods for nonlinear IEs could be mainly driven by the underlying testing mechanisms. 
Tests based on likelihoods should be closer to the criterion approaches.
Designing more specific tests for this task could also be targeted in future work.
Also the trees used in our analysis could be further optimized and examined. For example, weighting individual trees according gradient boosting-type procedures or applying alternative bootstrapping schemes could be beneficial. 
Another potentially impactful direction for future research is the investigation of lookahead strategies, including the implementation and optimization of lookahead algorithms that succeed in meta-analytic settings with a low number of studies and continuous variables.
With this, the trees could achieve better results for our second type of non-linear IE (\textit{Disc.-Time}).
Moreover, other tree-based approaches such as certain forms of prediction rule ensembles\autocite{fokkema2020Fitting} would also be interesting competitors.

\section{Practical guidance}
\label{Sec:guidance}
In conclusion, we provide several practical recommendations based on the results of our analysis.

When the number of studies $k$ is small, classical linear methods are the preferred choice. For very few studies these are also not optimal, particularly under substantial heterogeneity. For example, at $\tau^2 = 0.317$ and $k = 13$, they selected influential IEs only in 10--55\% of cases (see Figure~\ref{fig:combined_plot_extreme}).
However, trees frequently select no IEs at all in that case. 
The sample size at which trees become useful depends on the data structure and the number of covariates. In our simulation scenarios, they start selecting IEs from $k = 23$ onward (see detailed documentation of all scenarios in the Supplement, Figures S5 to S23).

When a strictly linear underlying model is plausible, classical linear variable selection methods are the natural choice. In such settings, there are no large differences between tests and criteria. 
An exception is univariate testing, which tests each variable individually and thereby has the advantage of being able to handle missing values mechanically but shows the expectable limitations, especially when covariates are correlated (see simulation study with synthetic covariate generation, Figure~S87)

To hedge somewhat against IEs that potentially actually have a non-linear structure, trees become more relevant.
For the simple non-linear IEs we considered, criterion-based selection remained superior in several settings (see Figure~\ref{fig:combined_plot_non_linear}).
However, especially for IEs of metric variables, the trees are competitive (not increasing Type~I error rates, and competitive Type~II error rates).
Thus, they represent a particular alternative, especially when even more complex non-linear IEs are expected and need to be accommodated within a classical inference model.
In the case of these actual model violations, the likelihood-based criterion procedures remain good, whereas the Wald-type tests should no longer be relied upon, as their Type~II error rates can increase substantially.

Among the tree-based approaches, stabilized ensembles generally perform best (lower Type~I and~II errors in most settings).
Among these, the random-effects variant (S-REmrt) performed better than the fixed-effect version (S-FEmrt), especially with respect to Type~I error rate (see, e.g., the plots in the second and fourth row of Figure~\ref{fig:combined_plot_extreme}).
As the selection rate grows with increasing $k$, the false selection rate for the S-REmrt increased less strongly than for the S-FEmrt, regardless of the heterogeneity present.

The stabilized approaches introduce an additional tuning parameter, $\lambda$, which governs the selection probability.
The parameter can be chosen according to the desired degree of conservatism and may also be used to counteract overly conservative behavior in settings with few studies by selecting lower values of $\lambda$. We deliberately refrain from reporting a universally optimal numerical value for $\lambda$, as the optimal choice depends on the specific analytical objective and characteristics of the data, including the number of studies $k$. Moreover, Type~I and Type~II error rates are not directly comparable across all simulation scenarios. As a general recommendation, we suggest selecting a central value of $\lambda$ in the range from 0.3 to 0.7, with $\lambda = 0.5$ serving as a reasonable default.
This also yields a reasonable trade-off between Type~I and~II error in our simulation, see Figure~\ref{fig:combined_plot_averaged_lambda}.

Alternatively, whenever feasible, we recommend inspecting the selection matrix $\boldsymbol{A}$ (Figure~\ref{fig:A_matrix}) rather than committing to a fixed $\lambda$ value. The empirical example demonstrates how examination of the selection matrices can substantially facilitate the interpretation of underlying structural patterns. 
Beyond its use for a full variable selection procedure, the stabilized trees, in combination with inspecting the selection matrix $\boldsymbol{A}$, can also serve as a complementary sensitivity analysis or as a pre-selection step prior to fitting a linear model.

\paragraph{Acknowledgements}
The authors gratefully acknowledge the computing time
provided on the Linux HPC cluster at Technical University
Dortmund (LiDO3), partially funded in the course of the
Large-Scale Equipment Initiative by the Deutsche
Forschungsgemeinschaft (DFG, German Research Foundation) as
project 271512359.

\paragraph{Author contribution} Conceptualization: J.-B.I.; Data curation: J.-B.I., P.L.; Formal analysis: J.-B.I., P.L.; Funding
acquisition: M.P.; Investigation: J.-B.I., P.L.; Methodology: J.-B.I.; Project administration: J.-B.I., M.P.;
Software: J.-B.I., P.L.; Supervision: M.P.; Visualization: J.-B.I., P.L.; Writing—original draft: J.-B.I.; Writing—review and editing: J.-B.I., M.P., P.L.

\paragraph{Conflict of interest}
The authors declare no potential conflict of interests.

\paragraph{Data availability statement}
The \texttt{R} script used for the analysis and simulations, together with additional tables and figures, is provided in the Supplementary Materials, published at \url{ https://doi.org/10.17877/TUDODATA-2026-3CDZSS}. The data of \citeauthor*{kimmoun2021Temporal}\autocite{kimmoun2021Temporal} is available at \url{https://osf.io/cxv5k/}. 

\paragraph{Financial disclosure}
All authors were supported by the German Research Foundation project, Grant no. 413270747. 

\paragraph{Declaration of generative AI and AI-assisted technologies in the writing process}
During the preparation of this manuscript, we used the ChatGPT 5.2 model from OpenAI and the Claude Sonnet 5 model from Anthropic for code and language edits, aiming to enhance readability. After using this tool/service, the authors reviewed and edited the content as needed and take full responsibility for the content of the manuscript.

\newpage
\printbibliography

\end{document}